\documentclass[doublecol]{epl2}
\usepackage{amsfonts}%
\usepackage{amsmath}%
\usepackage{amssymb}%
\usepackage{graphicx}
\usepackage{latexsym}

\newcommand{\bmath}{\begin{displaymath}}
\newcommand{\emath}{\end{displaymath}}
\newcommand{\bite}{\begin{itemize}}
\newcommand{\eite}{\end{itemize}}

\renewcommand{\P}{\mathcal{P}}

\newcommand{\bell}{\mathbf{\ell}}

\newcommand{\D}{\mathfrak{D}}

\newcommand{\one}{{1\!\!1}}
\newcommand{\half}{{\textstyle \frac{1}{2}}}
\newcommand{\E}{{\mathcal{E}}}

\newcommand{\bx}{\mathbf{x}}
\newcommand{\by}{\mathbf{y}}

\renewcommand{\O}{\mathcal{O}}
\newcommand{\R}{\mathcal{R}}

\newcommand{\bel}[1]{\begin{equation}\label{#1}}
\newcommand{\bal}[1]{\begin{eqnarray}\label{#1}}
\newcommand{\ee}{\end{equation}}
\newcommand{\ea}{\end{eqnarray}}

\newcommand{\Tr}{{\rm Tr}}
\newcommand{\tG}{{\tilde G}}
\newcommand{\tT}{{\tilde T}}
\newcommand{\tE}{{\tilde\E}}
\newcommand{\tphi}{{\tilde\phi}}

\newcommand{\fig}[1]{fig.~\ref{#1}}

\newcommand{\equ}[1]{eq.~(\ref{#1})}
\newcommand{\Equ}[1]{Eq.~(\ref{#1})}

\title{Irreducible Many-Body Casimir Energies of Intersecting Objects}
\shorttitle{Many-Body Casimir Energies} 

\author{M. Schaden} 

\institute{
   Department of Physics, Rutgers University, 101 Warren Street, Newark NJ 07102\\
  }

\abstract{
The vacuum energy of a bosonic field interacting locally with objects is decomposed into irreducible $N$-body parts. The irreducible $N$-body contribution to the vacuum energy is finite if the common intersection $O_1\cap O_2\dots \cap O_N$ of all $N$ objects $O_i,i=1,\dots, N$ is empty.  I prove that the perturbative expansion of the corresponding irreducible $N$-body spectral function $\tphi^{(N)}(\beta)$ for $\beta\sim 0$ vanishes to all orders even if some of the objects intersect. These irreducible spectral functions and their associated Casimir energies in principle can be computed numerically or approximated semiclassically without regularization or implicit knowledge of the spectrum. They are analytic in the parameters describing the relative orientation and position of the individual objects and remain finite when some, but not all, of the $N$ objects overlap. The Feynman-Kac theorem is used to compute Casimir energies of a massless scalar field with potential scattering and the finiteness of $N$-body Casimir energies is shown explicitly in this case. The irreducible $N$-body contributions to the vacuum energy of a massless scalar field with potential interactions is shown to be negative for an even- and positive for an odd- number of objects. Some simple examples are used to illustrate the analyticity of the $N$-body Casimir energy and its sign. A multiple scattering representation of the irreducible three-body Casimir energy is given. It remains finite when any two of the three objects overlap.}

\begin{document}

\maketitle

\section{Introduction}
The Casimir energy for two disjoint bodies is finite and may be estimated\cite{MS98,MSSS03,BMW06,KKN10,Graham01}. It can, in principle, be computed to arbitrary numerical precision\cite{EHGK01,GLM03,LM08}. For disjoint bodies, the multiple scattering representation of the interaction energy\cite{BB70,BD04,Klich06,EGJK08} thus solves many problems encountered in technological applications\cite{EHGK01,CMPS08}. We here develop an extension of this formalism and extract finite irreducible Casimir energies for more than two bodies that are not necessarily mutually disjoint. The analysis gives a new interpretation to finite parts of zero-point energies that could provide a framework for exploring gravitational effects due to vacuum energies\cite{SMPW07} and result in a more systematic approach to Casimir self-stresses for arbitrarily shaped bodies.

For clarity of presentation and to avoid infrared issues, we assume that the objects $\{O_i; i=1,\dots, N\}$ are all embedded in a large, but finite, connected Euclidean region $\D_\emptyset$ of dimension $d$. The thermodynamic limit $\D_\emptyset\rightarrow \R^d$ may be taken at the end. Formally, the vacuum energy $\E_{12\dots N}$ due to a massless bosonic field in the presence of $N$ objects can be decomposed into
\bel{decompose}
\E_{12\dots N}=\E_\emptyset+\sum_i\tE_i^{(1)}+\sum_{i<j}\tE_{ij}^{(2)}+\dots+\tE^{(N)}_{12\dots N}\ ,
\ee
where $\tE^{(k)}_{i_1\dots i_k}$ is the irreducible contribution to the vacuum energy that depends on \emph{all} $k$ objects $O_{i_1}\dots O_{i_k}$ in the domain $\D_\emptyset$. \Equ{decompose} would recursively define the irreducible $N$-body Casimir energy $\tE^{(N)}_{12\dots N}$ as an alternating sum of vacuum energies (see \equ{CasE}). The irreducible two-body Casimir energy $\tE^{(2)}_{12}$ for instance is
\bel{two-body}
\tE^{(2)}_{12}=\E_{12}-\E_{1}-\E_{2}+\E_\emptyset\ .
\ee
It is finite\cite{Klich06} for two disjoint objects. Below I show that the irreducible $N$-body Casimir energies $\tE^{(N)}_{12\dots N}$ are finite as long as the common intersection $O_1\cap O_2\dots \cap O_N$ of all $N$ objects is empty. For two objects this requires them to be disjoint, but three and more objects need not be mutually disjoint and the irreducible three-body contribution to the Casimir energy of a triangle, for instance, is finite. For a massless scalar field whose interaction with the objects is modeled by positive \emph{local} potentials, the irreducible $N$-body Casimir energies are shown to be negative for an even-, and positive for an odd-,  number of objects.

\section{Subtracted $N$-body Spectral Functions}
Some irreducible vacuum energies, such as self-energies $E^{(1)}$ may diverge when the associated objects overlap. \equ{decompose} thus is formal in the sense that it requires a high-frequency cutoff. However, spectral functions generally are finite and well defined even when the one-loop vacuum energies are not. They can be similarly decomposed into irreducible parts and we therefore relate the irreducible $N$-body Casimir energy $\tE^{(N)}$ to the corresponding irreducible $N$-body spectral function $\tphi^{(N)}(\beta)$,
\bel{casE}
\tE^{(N)}=-\frac{\hbar c}{\sqrt{8\pi}}\int_0^\infty \tphi^{(N)}(\beta)\frac{d\beta}{\beta^{3/2}}\ .
\ee
$\tphi^{(N)}(\beta)$ is constructed as follows. Let $\D_s$ represent the domain $\D_\emptyset$ with objects $\{O_j;j\in s\}$ embedded, $\D_{1\dots N}$ being the finite domain $\D_\emptyset$ with all $N$ objects included. Denote with $\mathfrak{P}(s)$ the power set of the elements of a set $s$ of finite cardinality $|s|\leq N$ with $\mathfrak{P}_N=\mathfrak{P}(\{1\dots N\})$.  Let $\phi_s(\beta)$ be the spectral function, or trace of the heat kernel $\mathfrak{K}_{\D_s}$, for the domain $\D_s$,
\bel{spfunc}
\phi_s(\beta)=\Tr\mathfrak{K}_{\D_s}(\beta)=\sum_{n\in\mathbb{N}} e^{-\beta\lambda_n(\D_s)/2}.
\ee
Here $\{\lambda_n(\D_s)>0,n\in \hbox{\text{I\hspace{-2pt}N}}$ is the spectrum of a bosonic field that vanishes on the boundary of $\D_\emptyset$ and whose interactions with the objects in $\D_s$ are \emph{local}. We assume the interaction of the field with the objects may be described by positive local potentials or take the form of (compatible) local boundary conditions.

\begin{figure}[hpbt]
\onefigure[angle=0,width=8.5cm]{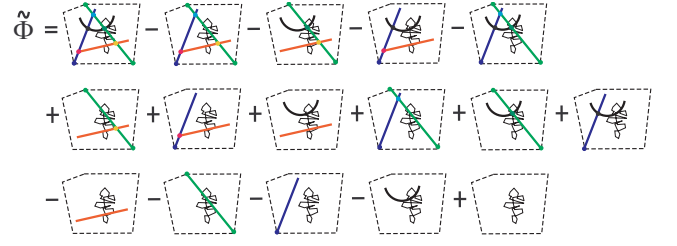}
\caption{\small The subtracted spectral function $\tphi^{(N)}(\beta)$ defined in  \equ{tphi} for a bounded two-dimensional domain $\D_\emptyset$ with four intersecting line segments as objects. Each pictograph represents the spectral function of the corresponding domain taken with the indicated sign. Various local features that contribute to the asymptotic expansion of each spectral function at high temperatures (small $\beta$) have been highlighted: lines of different color correspond to possibly different, but compatible, boundary conditions or local potentials. Since the intersections of line segments generally differ, each vertex is shown in a different color. The contribution to the asymptotic expansion from any particular local feature vanishes: the total signed number of times any particular line segment contributes is zero, as is the total signed number of times any particular vertex occurs. A random path that crosses three of the four segments is shown schematically. Its contribution to the Casimir energy of a scalar field in the Feynman-Kac path integral vanishes.}
\label{fourline}
\end{figure}

The irreducible spectral function $\tphi^{(N)}(\beta)$ of  \equ{casE} is the alternating sum of spectral functions $\phi_s(\beta)$ for the individual domains $D_s$,
\bel{tphi}
\tphi^{(N)}(\beta):=\sum_{s\in\mathfrak{P}_N} (-1)^{N-|s|}\phi_s(\beta)\ .
\ee
This is a special case of the geometrical subtraction procedure advocated in ref.\cite{Schaden09}. We will see that in this case the asymptotic expansion of the heat kernel vanishes to all orders if the common intersection of \emph{all} $N$ objects is empty. A pictorial representation of  \equ{tphi} for four line segments as objects in a bounded 2-dimensional Euclidean space is given in~\fig{fourline}.

To facilitate proving that the integral in  \equ{casE} is finite, we demand that the individual heat kernels are uniformly bounded by the free heat kernel of $\R^d$,
\bel{bound}
0<\mathfrak{K}_{\D_s}(\bx,\by;\beta)\leq K (2\pi\beta)^{-d/2} e^{-(\bx-\by)^2/(2\beta)}\ ,
\ee
for some finite $K>0$. For a scalar interacting with local positive potentials this is implied by the Feynman-Kac theorem\cite{FeynmanKac65}: the heat kernel is just the transition probability for Brownian motion and a (positive) potential \emph{reduces} it. Dirichlet boundary conditions in particular, may be imposed on a surface by killing any random path that crosses it. Because any condition is satisfied only by a reduced number of them, the bound also should hold for objects represented by any other local boundary condition. For the following it is sufficient that correlation functions vanish faster than any power of $\beta$ as $\beta\rightarrow 0$ for any \emph{finite} separation $(\bx-\by)^2>\delta^2>0$. One thus may be able to relax the uniform bound of \equ{bound} considerably.

We assume that the spectrum is discrete and positive semi-definite. $\phi_s(\beta)$ may be interpreted as a bosonic single particle partition function at inverse temperature $\beta$ and a positive spectrum is equivalent to the absence of tachyons in the causal local theory. The spectral functions $\phi_s(\beta)$ of \equ{tphi} in this case are positive and monotonically decreasing, approaching, at most, a finite positive constant for $\beta\sim\infty$.  Although we only treat a scalar bosonic system, the following also holds for electromagnetic fields in the absence of free charges.

In local field theories, the asymptotic expansion of $\phi_s(\beta)$ for small $\beta$  has the general form\cite{Greiner71,Gilkey84,Kirsten02,Vassilevich02},
\bel{highT}
\phi_s(\beta\sim 0)\sim \sum_{\nu=-d}^\infty (2\pi\beta)^{\nu/2} A^{(\nu)}_s+\O(e^{-\ell_\text{min}^2/(2\beta)})\ ,
\ee
where the Hadamard-Minakshisundaram-DeWitt-Seeley coefficients $A^{(\nu)}_s$ for the domain $\D_s$ have length-dimension $(-\nu)$. Note that if \equ{bound} holds, exponentially suppressed terms are associated with classical periodic paths of finite length $\ell_{min}$.  We decompose the heat kernel coefficients $A^{(\nu)}_s$ of \equ{highT} into parts arising from local features of the individual objects and their overlaps,
\bel{A}
A^{(\nu)}_s=\sum_{\tau\in\mathfrak{P}(s)} a^{(\nu)}_\tau \ ,
\ee
where the sum extends over all $(|s| !)$ sets in the power set $\mathfrak{P}(s)$ of the set $s$.  \equ{A} recursively defines reduced heat kernel coefficients $a^{(\nu)}_\tau $: the $a^{(\nu)}_\emptyset$ are the heat kernel coefficients associated with the Euclidean domain $\D_\emptyset$; the $a^{(\nu)}_{\{j\}}$ give their \emph{change} when object $j$ is inserted; the $a^{(\nu)}_{\{jk\}}$ account for further changes in the asymptotic heat kernel coefficients due to local overlaps of objects $j$ and $k$. Note that the $a^{(\nu)}_\tau$ are \emph{not} the heat kernel coefficients of the domain $\D_\tau$ -- they are their irreducible part only and arise from arbitrary short correlations near \emph{common} intersections of the objects in the set $\tau$.  $a^{(\nu)}_{\{jk\}}=0$ for two disjoint objects $j$ and $k$, if we assume (as implied by  \equ{bound}) that asymptotic correlations over finite distances $|\bx-\by|>\delta>0$ vanish faster than any power in $\beta$. Similarly, $a^{(\nu)}_{\{123\}}=0$ if $O_1\cap O_2\cap O_3=\emptyset$. $a^{(\nu)}_{\{123\}}=0$ vanishes even if the three objects are not mutually pairwise disjoint, $a^{(\nu)}_{\{jk\}}$ accounting for contributions to the asymptotic power series of the pairwise intersection $O_j\cap O_k$.

This argument may be extended to $N$ objects to imply that for local interactions the correction,
\bel{zero}
a^{(\nu)}_{\{1\dots N\}}=0, \text{  if   } O_1\cap\dots\cap O_N=\emptyset\ .
\ee
Note again that the condition in  \equ{zero} does not imply that the objects have to be mutually disjoint (except for $N=2$).  It then is a combinatoric matter to prove that for $\tau\subsetneq\{12\dots N\}$ the contribution of any non-zero $a^{(\nu)}_\tau$ to the asymptotic expansion of $\tphi^{(N)}(\beta)$ in  \equ{tphi} vanishes. Because the other $|s|-|\tau|$ objects may be selected from the remaining $N-|\tau|$ in any order, the number of times the set $\tau$ occurs as a subset of the sets in $\mathfrak{P}_N$ (with cardinality $|s|\ge|\tau|$) is the combination $\frac{(N-|\tau|)!}{(N-|s|)!(|s|-|\tau|)!}=\genfrac{(}{)}{0pt}{}{N-|\tau|}{N-|s|}$.  For $N>|\tau|$ the contribution to the asymptotic expansion of $\tphi^{(N)}(\beta)$ in \equ{tphi} proportional to $a^{(\nu)}_\tau$ then is
\bel{astot}
(2\pi\beta)^{\nu/2} a^{(\nu)}_\tau \sum_{|s|=|\tau|}^{N} (-1)^{N-|s|}\genfrac{(}{)}{0pt}{}{N-|\tau|}{N-|s|} =0.
\ee
When $N$ objects have no common intersection, the asymptotic expansion of the irreducible $N$-body spectral function $\tphi^{(N)}(\beta)$ 
thus has the form,
\bel{astphi}
\tphi^{(N)}(\beta\sim 0)\sim \O(e^{-\ell_\text{min}^2/(2\beta)})\ ,
\ee
and vanishes faster than any power of $\beta$. This may be explicitly verified in examples like that shown in \fig{fourline}, noting that contributions to the asymptotic expansion proportional to the volume, surfaces, corners, curvatures etc\dots, all cancel. Together with the fact that the spectral functions $\phi_s(\beta)$ decay monotonically and remain bounded for large $\beta$, the asymptotic behavior of  \equ{astphi} implies that the Casimir energy given by the integral in  \equ{casE} is \emph{finite}.

The subtraction procedure allows one to formally interpret $\tE^{(N)}$ as the alternating sum of vacuum energies $\E_s$ associated with the domains $\D_s$,
\bel{CasE}
\tE^{(N)}=\sum_{s\in \mathfrak{P}_N}(-1)^{N-|s|} \E_s\ .
\ee
The sum on the right side of  \equ{CasE} requires some regularization to be meaningful but, if this procedure does not explicitly depend on the specific domain $\D_s$ (for instance by regularizing the proper time integrals), the previous considerations show that the irreducible $N$-body contribution $\tE^{(N)}$ remains well defined as the regularization is removed. The absence of a power series in the asymptotic expansion of $\tphi^{(N)}(\beta\sim 0)$ also explains why a semi-classical approach based on classical periodic orbits tends to approximate Casimir energies fairly well\cite{MS98,MSSS03,BMW06,FKKLM09,Schaden09a,Schaden10}: it reproduces the leading exponentially suppressed terms of the asymptotic expansion.

\section{Massless Scalar Field with Local Potential Interactions}
The subtraction procedure we have just outlined is particularly transparent for a massless scalar field in a bounded Euclidean space $\D_\emptyset$ whose interaction with the objects is described by a local (positive) potential $V=\sum_i V_i$. It allows for an alternative proof in this case and provides an additional insight into the sign of the irreducible $N$-body contribution. Using the world-line approach of \cite{GLM03} for potential scattering, the Feynman-Kac theorem\cite{FeynmanKac65} generally states that,
\bel{support}
\phi_s(\beta)=\int_{\D_\emptyset} \frac{d\bx}{(2\pi \beta)^{d/2}} \P_{\D_s}[\bell_\beta(\bx)]\ ,
\ee
where $\P_{\D_s}[\bell_\beta(\bx)]$ is the probability for a standard Brownian bridge\footnote{A standard Brownian bridge $\bell_\beta(\bx)=\{\bx+\sqrt{\beta}(\mathbf{W}(t)-t\mathbf{W}(1)); 0\leq t\leq 1\}$ is generated by a standard $d$-dimensional Wiener process with stationary and independent increments for which $\mathbf{W}(t>0)$ is normally distributed with variance $t d$ and vanishing average.}, $\bell_\beta(\bx)=\{\bx_t, 0\leq t\leq\beta; \bx_0=\bx_t=\bx\}$, that starts at $\bx$ and returns to $\bx$ after "time" $\beta$, to not exit $\D_s$ and survive its encounters with the objects. The survival probability of any particular Brownian bridge in $\D_s$ is given by $p_s(\bell_\beta(\bx))=\exp[-\int_0^\beta V_s(\bx_t) dt]$, where $V_s$ is the sum of local potentials representing the objects in $\D_s$. Dirichlet boundary conditions in particular may be imposed by setting $p_s=0$ for a loop that crosses the surface of the object and $p_s=1$ if it does not.

The contribution to $\tphi^{(N)}(\beta)$ of a loop $\ell^{(\tau)}_\beta(\bx)$ that remains within $\D_\emptyset$ and encounters all objects of $\tau\subsetneq\{1\dots N\}$ and no others is,
\bel{loopcont}
\sum_{{\genfrac{}{}{0pt}{}{s\in \mathfrak{P}_N}{s\cap\tau=\gamma}}}\hspace{-.5em}p_\gamma(-1)^{N-|s|}=\sum_{\gamma\in\mathfrak{P}(\tau)}\hspace{-.6em}p_\gamma\hspace{-1em} \sum_{s=|\gamma|}^{N-|\tau|+|\gamma|}\hspace{-1em}(-1)^{N-s}\genfrac{(}{)}{0pt}{}{N-|\tau|}{s-|\gamma|}=0\ ,
\ee
The contribution vanishes independent of the survival probabilities $p_\gamma$. Only loops that touch all $N$ objects, ($\tau=\{1,\dots,N\}$) contribute to the alternating sum in  \equ{tphi} and we have that
\bel{pathtphi}
\tphi^{(N)}(\beta)=(-1)^N\int_{\D_\emptyset} \frac{d\bx}{(2\pi \beta)^{d/2}} \tilde\P^{(N)}[\bx;\beta]\ ,
\ee
where $\tilde\P^{(N)}[\bx;\beta]$ is the probability that a standard Brownian bridge starting at $\bx$ and returning to $\bx$ after "time" $\beta$ does not exit $\D_\emptyset$ and is killed by \emph{every} one of the $N$ objects. Composing the probabilities one observes that a Brownian bridge is killed by every one of $N$ objects with probability,
\bel{killprob}
p(\text{killed by every one of $N$ objects})=\sum_{\gamma\in\mathfrak{P}_N}(-1)^{|\gamma|}p_\gamma\ .
\ee
\equ{killprob} is the extension to $N$ objects of the statement,
\bel{composeprob}
\begin{array}{l}
p(\text{killed by $O_1$ \emph{and} killed by $O_2$})=\\
\ =(1-p_1)+(1-p_2)-(1-p_{12})=p_\emptyset-p_1-p_2+p_{12}.
\end{array}\nonumber
\ee
Note that Eqs.~(\ref{pathtphi})~and~(\ref{killprob}) do not require survival probabilities to be independent, $p_{12}=p_1p_2$ -- which is the case only for potentials that do not overlap (and therefore corresponds to mutually disjoint objects). Since $\tilde\P^{(N)}$ is a positive probability, the factor of $(-1)^N$ in \equ{pathtphi} determines the sign of $\tphi^{(N)}(\beta)$. [For Dirichlet boundary conditions on the objects, it arises because paths that touch all $N$ objects contribute only to $\phi_\emptyset(\beta)$.]  For scalar fields interacting by local potentials with $N$ objects that have no common intersection, the irreducible $N$-body Casimir energy is finite and its sign is given by,
\bel{signE}
(-1)^N\tE^{(N)}<0.
\ee
It is remarkable that the sign of $\tE^{(N)}$ depends only on the number of objects and the fact that the interaction with the scalar field is described by local potentials. The irreducible scalar two-body Casimir energy in particular, in this case is negative independent of mirror symmetry\cite{Klich06}. This also holds for Dirichlet boundary conditions. But \equ{signE} does not hold for boundary conditions like Neumann's,  that are not described by potentials\cite{Boyer74}. Also, the sign of the $N$-body Casimir energy does not of itself determine whether Casimir \emph{forces} are attractive or repulsive. The subtraction clearly exhibits the irreducible part of the vacuum energy computed in \equ{CasE}, but it is important to correctly interpret this energy. The finite $N$-body Casimir energy obtained here is the irreducible $N$-body \emph{correction} to the vacuum energy that remains when \emph{all} $M$-body vacuum energies with $0\leq M<N$ have been accounted for. The latter may themselves be finite but very often are not, and the sign of $\tE^{(N)}$ given in  \equ{signE} is that of the irreducible $N$-body part only, which, in general, does not coincide with the work needed to assemble the $N$ objects.

For a scalar field,  \equ{pathtphi} interprets $\tphi^{(N)}(\beta)$ as a probability for a random walk that satisfies certain geometric conditions. Since they have to touch $N$ objects that have no common intersection, Brownian bridges that contribute in \equ{pathtphi} necessarily are of \emph{finite} length. The probability $\tilde\P^{(N)}[\bx;\beta]$ thus is bounded from above by the shortest closed classical path of length $\ell_\text{min}$ that just touches \emph{all} objects,
\bel{boundary}
0\leq\tilde\P^{(N)}[\bx;\beta]\leq e^{-\ell^2_{min}/(2\beta)}\ .
\ee
For a domain $\D_\emptyset$ of finite volume, the bound of \equ{boundary} implies that the asymptotic power series in $\beta$ of $\tphi^{(N)}(\beta\sim 0)$ vanishes to all orders in this case, as we have previously argued more generally.

\section{Examples}
Consider the example of a scalar field in $\R^{d}$ satisfying Dirichlet boundary conditions on $(d+1)$ intersecting, $(d-1)$-dimensional hyper-planes. In this case $\tE^{(d+1)}$ indeed is the work required to adiabatically move the last hyperplane into position from infinity: $\tE^{(d+1)}$ vanishes as the volume enclosed by the hyper-planes becomes infinite and depends continuously on their position. These are simple consequences of the smoothness and continuity of the probability that Brownian bridges cross all of them in time $\beta$.  \equ{CasE} implies that (infinite) hyper-planes forming a simplex, such as a triangle($d=2$) or a pyramid($d=3$), tend to repel (triangle) in even and to attract (pyramid) in odd dimensional spaces. Contributions to Casimir energies from interior modes of domains with generalized reflection symmetries have been computed analytically as well as numerically \cite{Dowker06,Almedov05,Milton10}. None of these obtain only the finite irreducible part of the three-body Casimir energy and the results are somewhat ambiguous because corner divergences are subtracted in different ways. The world-line method\cite{GLM03,Schaden09,Schaden09a} outlined above could provide fairly accurate numerical estimates for the irreducible $N$-body part of the Casimir energy of a scalar field with Dirichlet boundary conditions on any generic set of $N$ intersecting hyper-planes without restricting to contributions from interior modes. The irreducible three-body Casimir energies of some weakly interacting intersecting objects are explicitly found to be positive and finite \cite{future}.

The analytically tractable Casimir energy of $2^{d}$ pairwise parallel $(d-1)$-dimensional hyper-planes forming a multi-dimensional tic-tac-toe-like pattern in $\R^{d}$ that encloses an inner hyper-rectangle with dimensions $\ell_1\times\dots\times \ell_d$ was previously considered in\cite{SvaiterSvaiter92}, without separation of the analytic irreducible $2^d$-body contribution. The irreducible $2^d$-body Casimir energy $\tE^{2^d}_\text{rect.}$ for a scalar field satisfying Dirichlet conditions on all the hyper-planes in fact has the simple form\cite{future}
\bel{hrectangle}
\tE^{2^{d}}_\text{rect.}=-\frac{\hbar c \Gamma[(d+1)/2]}{4\pi^{(d+1)/2}}\sum_{n_1=1}^\infty\dots\sum_{n_d=1}^\infty\frac{{\cal V}_\text{rect.}}{L^{d+1}(\mathbf{n})},
\ee
where  ${\cal V}_\text{rect.}=\prod_{j=1}^d \ell_j$ is the volume of the hyper-rectangle and $L(\mathbf{n})=\sqrt{\sum^d_{j=1} n_j^2 \ell_j^2}$ is half the length of a classical periodic orbit in its \emph{interior} that reflects $n_j$ times off the $j$-th pair of parallel hyper-planes. Only classical periodic orbits that touch all hyper-planes contribute to $\tE^{2^{d}}_\text{rect.}$.  Consistent with the previous results,  $\tE^{2^{d}}_\text{rect.}$ is negative and finite and remains so in the limit in which one or more dimensions of the rectangle vanish and up to $(d-1)$ pairs of hyper-planes coincide. Note that $\tE^{2^{d}}_\text{rect.}(\ell_k\rightarrow 0)=\half\tE^{2^{d-1}}_\text{rect.}$. When \emph{any} dimension of the rectangle becomes large $\tE^{2^{d}}_\text{rect.}$ vanishes. As mentioned previously this analyticity in the shape and dimensions of the objects is expected in the world-line description and is one of the more interesting characteristics of the irreducible $N$-body Casimir energies defined by  \equ{casE} and  \equ{tphi}. One might ask whether  \equ{hrectangle} can be given a meaning that does not involve divergent irreducible contributions involving less than $2^d$-bodies. In fact $\tE^{2^{d}}_\text{rect.}$ determines higher order derivatives of the vacuum energy in much the same manner as the original Casimir energy determines the force between two plates:
\bel{interpret}
\partial_{\ell_1}\partial_{\ell_2}\dots\partial_{\ell_d}\E_\text{rect.}=\partial_{\ell_1}\partial_{\ell_2}\dots\partial_{\ell_d}\E^{2^d}_\text{rect.}\ ,
\ee
since all irreducible contributions with less than $2^d$ bodies do not depend on all dimensions of the rectangle. For $d=2$ pairs of parallel plates \equ{interpret} implies that the irreducible four-body contribution to the vacuum energy fully describes certain stability derivatives of the vacuum energy. Other irreducible finite $N$-body Casimir energies also suffice to describe, generally more complicated, higher order derivatives of the vacuum energy. Finite irreducible $N$-body Casimir energies do not depend sensitively on the (quantum) description of the intersecting objects at high scattering energy and thus are reliably estimated by low-energy effective models that describe the interaction with the objects in terms of potentials or boundary conditions.

\begin{figure}[hpbt]
\onefigure[angle=0,width=8.5cm]{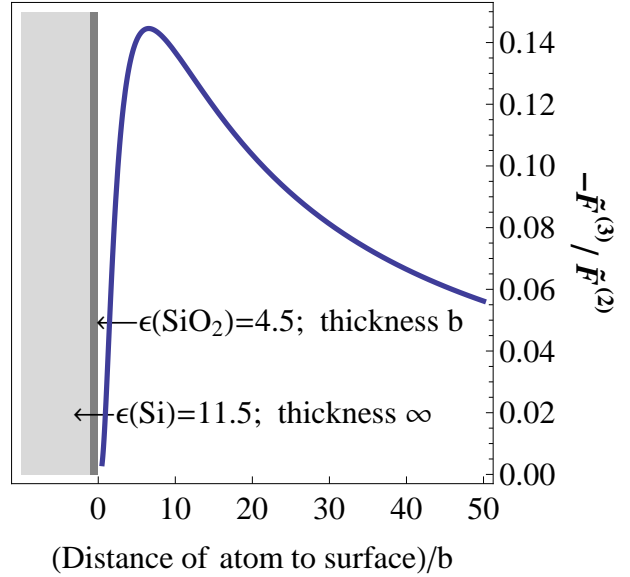}
\caption{\small Ratio of the irreducible three-body Casimir force to the sum of irreducible two-body Casimir forces on an atom near a Si/SiO$_2$ bilayer. The distance of the atom from the surface is measured in units of the thickness, $b$, of the SiO$_2$ layer.}
\label{bilayer}
\end{figure}

An estimate of the magnitude of irreducible three-body Casimir forces is provided by the Casimir-Polder force on a polarizable atom due to a bilayer. Using the results of \cite{Spruch95}, \fig{bilayer} compares the irreducible electromagnetic three-body Casimir force on an atom with the sum of irreducible two-body Casimir forces for a Si/SiO$_2$ bilayer. It is well known that Casimir forces are not additive and the three-body correction in this example is not negligible. It \emph{diminishes} the overall attractive force by almost 15\% for atoms that are about 10 layer-thicknesses from the bilayer.

\section{Irreducible Casimir Energies in the Multiple Scattering Expansion}
To address more complicated geometries in the electromagnetic case, a representation of the irreducible Casimir energies in terms of one-body $T$-matrices is required\cite{EGJK08}.  For three bodies it may be obtained using the generating functional approach of\cite{Bordag85}. The irreducible three-body Casimir energy $\tE^{(3)}$ expressed in terms of the free-, 1-, 2- and 3-body Green's functions in the notation of \cite{CMPS08} is,
\bal{Gfunction}
\tE^{(3)}&=&{\textstyle \frac{i}{2\tau}}\Tr (\ln G_{123}-\ln G_{12}-\ln G_{23}-\ln G_{13}\nonumber\\
&&+\ln G_1+\ln G_2+\ln G_3-\ln G_\emptyset)\hspace{3em}\\
&&\hspace{-4em}={\textstyle \frac{-i}{2\tau}}\Tr (\ln \tG_1 \tG^{-1}_{123}\tG_{23}-\ln \tG_1\tG^{-1}_{12}\tG_2-\ln \tG_1\tG^{-1}_{13}\tG_3)\ ,\nonumber
\ea
where $G_\alpha=G_\emptyset\tG_\alpha$ is the Green's function for the domain $\D_\alpha$. The trace is over space and time, with $\tau$ here denoting the temporal extent. Using  $\tG^{-1}_{ij}=\tG_i^{-1}+\tG_j^{-1}-\one$ and $\tG^{-1}_{123}=\tG_1^{-1}+\tG_{23}^{-1}-\one$ with $\tG_i=\one-\tT_i$, the irreducible three-body Casimir energy of \equ{Gfunction} in terms of one-body scattering matrices $T_i$ finally is,
\bal{Tmatrix}
\tE^{(3)}&=&{\textstyle \frac{-i}{2\tau}}\Tr (\ln [\one-\tT_1(\one-\tG_{23})]+\ln[X_{12}]-\ln[X_{13}]\nonumber\\
&=&{\textstyle \frac{-i}{2\tau}}\Tr\ln \Big[\one-X_{12}\tT_1 \Big(\tT_2\tT_1\tT_3-\tG_2\tT_3 X_{23}\tT_2\nonumber\\
&&\qquad-\tG_3\tT_2 X_{32}\tT_3\Big)X_{13}\Big]\ .
\ea
Here $\tT_i= T_i G_\emptyset=(\one-\tG_i)$, with $G_\emptyset$ the Green's function for the domain $\D_\emptyset$ with no objects inserted. The operators $X_{ij}$ satisfy the integral equation,
\bel{def-X}
X_{ij}(\one-\tT_i\tT_j)=\one\ .
\ee
The expression in  \equ{Tmatrix} differs from that given in\cite{EGJK08} only in that all two-body interactions have been subtracted. As our previous considerations show, the irreducible three-body Casimir energy given in\equ{Tmatrix} is continuous in the position of the three objects and remains finite when two overlap pairwise even as the corresponding (irreducible) two-body contribution to the vacuum energy diverges. Every term in  \equ{Tmatrix} requires scattering off all three objects and is individually finite. We explicitly calculated\cite{future} the three-body correction of \equ{Tmatrix} to the Casimir energy of three semi-transparent parallel plates\cite{future}. The result is symmetric under the exchange symmetry and remains finite when any two of the three plates coincide.  Writing irreducible $N$-body Casimir energies in terms of scattering matrices unambiguously defines them for any local field theory and in particular for the electromagnetic case.

\acknowledgments
I would like to thank S.A.~Fulling K.A.~Milton and K.V.~Shajesh for helpful discussions and improvements to the
manuscript. This work was supported by the National Science Foundation with Grant no.~PHY0555580.


\begin{thebibliography}{0}

\bibitem{MS98}
\Name{Schaden M., \and Spruch L.}
\REVIEW{Phys. Rev. A}{58}{1998}{935};\REVIEW{Phys. Rev. Lett.}{84}{2000}{459};\REVIEW{Phys. Rev. A}{65}{2002}{022108}.

\bibitem{MSSS03}
\Name{Mazzitelli F.D.,Sanchez M.J., Scoccola N.N. \and von Stecher J.}
\REVIEW{Phys. Rev. A}{67}{2003}{013807}.

\bibitem{BMW06}
\Name{Bulgac A.,Magierski P. \and Wirzba A.}
\REVIEW{Phys. Rev. D}{73}{2006}{025007}.   

\bibitem{KKN10}
\Name{Kabat D., Karabali D. \and Nair V.P.}
\REVIEW{Phys. Rev. D}{81}{2010}{125013}.  


\bibitem{Graham01}
\Name{Graham N., Jaffe R.L., Khemani V., Quandt M., Scandurra M. \and Weigel H.}
\REVIEW{Nucl. Phys. B}{645}{2002}{49};
\Name{Graham N., Jaffe R.L., Quandt M., Schr\"{o}der O. \and Weigel H.}
\REVIEW{Nucl. Phys. B}{677}{2004}{379};
\Name{Schr{\"o}der O., Scardicchio A. \and Jaffe R.L.}
\REVIEW{Phys. Rev. A}{72}{2005}{012105}.  

\bibitem{EHGK01}
\Name{Emig T., Hanke A., Golestanian R. \and Kardar M.}
\REVIEW{Phys. Rev. Lett.}{87}{2001}{260402};
\Name{B{\"u}scher R. \and Emig T.}
\REVIEW{Phys. Rev. Lett.}{94}{2005}{133901};
\Name{Emig T., Graham N., Jaffe R.L. \and Kardar M.}
\REVIEW{Phys. Rev. Lett.}{99}{2007}{170403};
\Name{Reid M.T.H., Rodriguez A.W., White J. \and Johnson S.G.}
\REVIEW{Phys. Rev. Lett.}{103}{2009}{040401};
\Name{Graham N., Shpunt A., Emig T., Rahi S.J., Jaffe R.L. \and Kardar M.}
\REVIEW{Phys. Rev. D}{81}{2010}{061701};  
\Name{Maghrebi M.F., Rahi S.J., Emig T., Graham N., Jaffe R.L. \and Kardar M.}
\REVIEW{arXiv}{quant-ph}{1010}{3223}.

\bibitem{GLM03}
\Name{Gies H. \and Langfeld K.}
\REVIEW{Int. J. Mod. Phys. A}{17}{2002}{966} 
\Name{Gies H., Langfeld K. \and Moyaerts L.}
\REVIEW{JHEP}{0306}{2003}{018} 
\Name{Gies H. \and Klingm{\"u}ller K.}
\REVIEW{Phys. Rev. Lett.}{96}{2006}{220401}   
\Name{Weber A.\and Gies H.}
\REVIEW{Phys. Rev. Lett.}{105}{2010}{040403};  
\Name{Gies H. \and Klingm{\"u}ller K.}
\REVIEW{Phys. Rev. Lett.}{97}{2006}{220405}.  

\bibitem{LM08}
\Name{Lambrecht A. \and Marachevsky V.N.}
\REVIEW{Phys. Rev. Lett.}{101}{2008}{160403};
\REVIEW{Int. J. Mod. Phys. A}{24}{2009}{1789};
\Name{Chiu H.-C., Klimchitskaya G.L., Marachevsky V.N., Mostepanenko V.M. \and Mohideen U.}
\REVIEW{Phys. Rev. B}{81}{2010}{115417}.

\bibitem{BB70} 
\Name{Balian R.B. \and Bloch C.}
\REVIEW{Ann. Phys. (NY)}{60}{1970}{401};
\REVIEW{ibid}{63}{1971}{592};
\REVIEW{ibid}{64}{1971}{271};
\REVIEW{Errata}{84}{1974}{559};
\REVIEW{ibid}{69}{1972}{76};
\REVIEW{ibid}{85}{1974}{514}.

\bibitem{BD04}
\Name{Balian R. \and Duplantier B.}
\REVIEW{Annals Phys. (NY)}{104}{1977}{300};
\REVIEW{ibid}{112}{1978}{165};
\Name{Balian R. \and Duplantier B.}
  \Book{Recent Developments in Gravitational Physics: Institute of Physics Conference Series}
  \Editor{A.Ciufiolini et al.}
  \Vol{176}
  \Publ{Taylor \& Francis, Boca Raton}
  \Year{2004}
  \Page{1, arXiv: quant-ph/0408124}.

\bibitem{Klich06}
\Name{Kenneth O. \and Klich I.}
\REVIEW{Phys. Rev. Lett.}{97}{2006}{160401};
\Name{Bachas C.P.} 
\REVIEW{J. Phys. A}{40}{2007}{9089}.    

\bibitem{EGJK08}
\Name{Emig T., Graham N., Jaffe  R.L. \and Kardar M.}
\REVIEW{Phys. Rev. D}{77}{2008}{025005} see eq.~(III.27);  
\Name{Emig T.\and Jaffe R.L.}
\REVIEW{J. Phys. A}{41}{2008}{164001}.

\bibitem{CMPS08}
\Name{Cavero-Pelaez I., Milton K.A., Parashar P. \and Shajesh  K.V.}
\REVIEW{Phys. Rev. D}{78}{2008}{065018}. 

\bibitem{SMPW07}
\Name{Fulling S.A., Milton K.A., Parashar P., Romeo A., Shajesh K.V. \and Wagner J.}
\REVIEW{Phys. Rev. D}{76}{2007}{25004};  
\Name{Estrada R., Fulling S.A., Liu Z., Kaplan L., Kirsten K. \and Milton  K.A.}
\REVIEW{J. Phys. A}{41}{2008}{164055}.

\bibitem{Schaden09}
\Name{Schaden M.}
\REVIEW{Phys. Rev. Lett.}{102}{2009}{060402}.

\bibitem{FeynmanKac65}
\Name{Feynman R.P. \and Hibbs A.R.}
\Book{Quantum Mechanics and Path Integrals}
\Publ{McGraw-Hill, New York}
  \Year{1965};
\Name{Kac M.}
\REVIEW{Amer. Math. Monthly}{73}{1966}{1}.

\bibitem {Greiner71}
\Name{Greiner P.}
\REVIEW{Arch. Rat. Mech. Anal.}{41}{1971}{163}.

\bibitem {Gilkey84} P. B. Gilkey,
\Name{Gilkey P.B.}
\Book{Invariance Theory, the Heat Equation, and the Atiyah-Singer Index Theorem}
\Publ{Publish or Perish, Wilmington}
\Year{1984} and
\Publ{CRC Press, Boca Raton}
\Year{1995}.

\bibitem {Kirsten02}
\Name{Kirsten K.}
\Book{Spectral functions in Mathematics and Physics}
\Publ{Chapman \& Hall/CRC Press, Boca Raton}
\Year{2002}.

\bibitem{Vassilevich02}
\Name{Vassilevich D.V.}
\REVIEW{Physics Rep.}{388}{2003}{279}.

\bibitem{Schaden09a}
\Name{Schaden M.}{Phys. Rev. A}{79}{2009}{052105}.

\bibitem{Schaden10}
\Name{Schaden M.}
\REVIEW{Phys. Rev. A}{73}{2006}{042102};
\REVIEW{Phys. Rev. A}{82}{2010}{022113}.

\bibitem{FKKLM09}
\Name{Fulling S.A., Kaplan L.,Kirsten K., Liu Z.H. \and Milton K.A.}
\REVIEW{J. Phys. A}{42}{2009}{155402}.   

\bibitem{Dowker06}
\Name{Dowker J.S.}
\REVIEW{Class. Quant. Grav.}{23}{2006}{2771}.

\bibitem{Almedov05}
\Name{Ahmedov H. \and Duru I.H.}
\REVIEW{J. Math. Phys.}{46}{2005}{022303};\REVIEW{ibid}{46}{2005}{022304};\REVIEW{Phys. Atom. Nuclei}{68}{2005}{1621};
\Name{Ahmedov H.}
\REVIEW{J. Phys. A}{40}{2007}{10611}.

\bibitem{Milton10}
\Name{Abalo E.K., Milton K.A. \and Kaplan L.} preprint [arXiv:1008.4778v1].

\bibitem{SvaiterSvaiter92}
\Name{Svaiter N.F. \and Svaiter B.F.}
\REVIEW{J. Phys. A}{25}{1992}{979}.

\bibitem{future}
\Name{Shajesh K.V. \and Schaden M.} in preparation.

\bibitem{Bordag85}
\Name{Bordag M., Robaschik D. \and Wieczorek E.}
\REVIEW{Ann. Phys. (N.Y.)}{165}{1985}{192};
\Name{Robaschik D., Scharnhorst K. \and Wieczorek E.}
\REVIEW{Ann. Phys. (N.Y.)}{174}{1987}{401};
\Name{Bordag M., Hennig D. \and Robaschik D.}
\REVIEW{J. Phys. A}{25}{1992}{4483};
\Name{Bordag M.}
\REVIEW{Phys. Rev. D}{73}{2006}{125018}.

\bibitem{Boyer74}
\Name{Boyer, T.H.}
\REVIEW{Phys. Rev. A}{9}{1974}{2078};
\Name{Hushwater, V.}
\REVIEW{Am. J. Phys.}{65}{1997}{381};
\Name{Schaden, M. \and Spruch, L.}
\REVIEW{Phys. Rev. A}{58}{1998}{935}.

\bibitem{Spruch95}
\Name{Zhou F, \and Spruch L.}
\REVIEW{Phys. Rev. A}{52}{1995}{297};
\Name{Salem R., Japha Y., Chabé J., Hadad B., Keil M., Milton K.A. \and Folman R.}
\REVIEW{New J. Phys.}{12}{2010}{023039}, Appendix~A.
\end{thebibliography}
\end{document}